\documentclass[universe,article,accept,pdftex,moreauthors]{Definitions/mdpi}
\firstpage{1} 
\pubvolume{1}
\issuenum{1}
\articlenumber{0}
\pubyear{2024}
\copyrightyear{2024}
\datereceived{} 
\dateaccepted{} 
\datepublished{} 
\hreflink{https://doi.org/}

\Title{
Radio observations as a probe of cosmic web magnetism }

\TitleCitation{
Radio observations as a probe of cosmic web magnetism }

\Author{Ettore Carretti$^{1,}$*\orcidA{}, and Franco Vazza$^{2,1}$\orcidB{}}

\AuthorNames{Ettore Carretti and Franco Vazza}

\address{%
$^{1}$ \quad INAF - Istituto di Radioastronomia, Via Gobetti 101, 40129, Bologna, Italy
\\
$^{2}$ \quad Dipartimento di Fisica e Astronomia, Universit\'a di Bologna, via Gobetti 93/2, 40122 Bologna, Italy}

\corres{Correspondence: carretti@ira.inaf.it (E.C.)}

\abstract{The Universe's magnetogenesis can be investigated with radio observations of cosmic filaments, where the information on the initial magnetic field seeds is expected to be preserved in time. In this work, we update the comparison between  recent observational results in filaments with the predictions from recent cosmological simulations to check whether one of them is favoured.  The radio probes we use are the rotation  measure (RM) of filaments as a function of the redshift ($z$), stacking  of synchrotron emission from filaments, and the RM radial profile away from galaxy groups.  The first two probes  favour the presence of a dominant primordial magnetic field component and disfavour a sole astrophysical scenario, the third probe does not yet give an unambiguous outcome. We also estimate the average field strength in filaments. Independently of the scenario and the shape of the astrophysical component RM, it is in the range 10--60 nG at $z=0$, while, when restricted to the model that gives the best match to the simulations, it gives $43\pm 7$ nG,  with an astrophysical component RM rapidly decreasing with the redshift. }

\keyword{magnetic fields; intergalactic medium; large scale structure of Universe; methods: statistical} 

\begin{document}


\section{Introduction}

 {The} 
origin of extragalactic magnetic fields is a key element in studying the early Universe and, currently, can be investigated with complementary observables  {(e.g.,} 
\citep{2016RPPh...79g6901S,2019JCAP...11..028P,2019PhRvL.123b1301J,2021Univ....7..223A,2021RPPh...84g4901V, 2021MNRAS.505.5038A,2021MNRAS.500.5350V,2024A&A...683A..25V}). There are two major groups of models to explain  {magnetogenesis:} 
\endnote{The magnetogenesis is the generation of the seed fields of large-scale magnetic fields in the Universe.} 
primordial scenarios, where the fields were produced early, during cosmic inflation or in phase transitions before recombination (e.g., \citep{1988PhRvD..37.2743T, 1994RPPh...57..325K, 2019JCAP...11..028P}), and
astrophysical scenarios, where the fields were generated at later times, after or during the formation of galaxies by star formation, jets and outflows from active galactic nuclei (AGNs), or batteries at shocks or at ionization fronts \citep[][]{1994RPPh...57..325K, 2006MNRAS.370..319B, 2009MNRAS.392.1008D, va25a}.

Cosmic filaments represent a sweet spot for studying cosmic magnetism and its origin, since the magnetic field there is expected to be little processed by cosmic evolution, unlike in galaxies or clusters of galaxies \citep{2014ApJ...797..133C,2017CQGra..34w4001V}), thus preserving the information on the primeval field origin (e.g., \citep{2021MNRAS.505.4178V, 2023MNRAS.518.2273C, 2025A&A...693A.208C}). At the same time, the strength of magnetic fields is expected to be higher in filaments than in voids, making their detection easier \citep{2010Sci...328...73N,2017CQGra..34w4001V, 2022ApJ...929..127M}.

Observations in the radio band are powerful probes for investigating the magnetization of the Universe, down to a range of cosmic overdensities at which different magnetogenesis scenarios can be robustly discriminated. The present sensitivity of modern radio surveys, combined with the sophisticated modeling of numerical simulations, makes it possible to derive accurate constraints on the properties of primordial magnetic fields. These constraints are significantly better than the best derived from the analysis of the cosmic microwave background (CMB), for a class of models (e.g., \citep{2024arXiv241214825N} (for a recent review)).

In this work, we review the current constraints on the magnetization of the Universe on large scales with radio observations (Section \ref{sec:radio}). We show how the combination with modern cosmological simulations allows us already to holistically test magnetogenesis scenarios (Section \ref{sec:sim}). We perform improved analyses to check which magnetogenesis scenario is favored, by using the comparison between recent simulations and recent observations: rotation measure in filaments as a function of the redshift; stacking of synchrotron emission in filaments; and rotation measure radial profile moving away from galaxy groups. These comparisons allow us to select a currently best working scenario, which stems from the combination of astrophysical and primordial magnetic fields (Section \ref{sec:analysis}). Finally, we discuss our results in Section \ref{sec:discussion} and draw our conclusions in Section \ref{sec:conc}.

\section{Current Constraints from Observations}
\label{sec:radio}
\subsection{Synchrotron Emission}
\label{sync}
Synchrotron radiation is emitted in the radio domain by ultra-relativistic electrons spiraling when immersed in magnetic fields, and it is one of the most powerful detection tools for investigating cosmic magnetic fields (e.g., \citep{1994RPPh...57..325K}). In the cosmic web, it is also known as the synchrotron cosmic web.

\textls[-15]{A useful formula (derived by Ref. \citep{hb07} in the context of radio relic emission in clusters of galaxies) predicts the synchrotron emission in the downstream of cosmic shocks by assuming the production of power-law distributions of relativistic electrons injected from the thermal pool of particles, through the process of diffusive shock acceleration (DSA, e.g., \cite[][]{be78,dv81,kj07}):} 
\vspace{-6pt}
\begin{eqnarray}
P_{\nu}=6.4 \times 10^{34}{[\rm erg/s/Hz]} \times {\xi_e}(\mathcal{M}) \frac{A}{\rm Mpc^2} ~\frac{n_e}{10^{-4}\rm cm^{-3}} \nonumber\\ 
~(\frac{T}{7 \rm keV})^{3/2} (\frac{\nu}{\rm GHz})^{-s/2} \times \frac {B^{1+s/2}}{B_{\rm CMB}^2+B^2} \frac{1}{(1+z)^2},
\label{eq:hb}
\end{eqnarray}
 {where} 
$A$ is the shock surface, $T$ is the upstream temperature, $\mathcal{M}$ is the shock Mach number, $s$ is the energy spectrum slope of the accelerated particles (which is $s=2(\mathcal{M}^2+1)/(\mathcal{M}^2-1)$ in DSA), $B$ is the magnetic field (in $\mu G$), and $B_{\rm CMB}$ is the CMB-equivalent magnetic field. $\nu$ is the observing frequency, and $\xi_e(\mathcal{M})$ is the acceleration efficiency of electrons. 
Numerical simulations generally assumed $\xi_e(\mathcal{M}) \approx 10^{-2}$ for strong shocks ($\mathcal{M} \gg 10$) and much less ($\leq 10^{-5}-10^{-4}$) for very weak shocks, but in general, $\xi_e$ is a poorly known quantity from basic principles (e.g., \citep{ka07,guo14,Bykov19,gupta24}), because directing the injection of electrons from the thermal pool is problematic in DSA. 

In this work, and in line with the literature, we will use Equation (\ref{eq:hb}) to estimate the level of radio emission by cosmic shocks even outside clusters of galaxies, with the important caveat that this assumes that DSA can operate even at lower densities and magnetization levels than in the intracluster medium. 
This is partially supported by particle-in-cell (PIC) simulations, which found that the topology of the upstream
magnetic field is more important than its initial value, for the onset of DSA in non-relativistic shocks. The rapid development of the Weibel instability might amplify magnetic fields and enable the formation of a shock precursor, leading to DSA in quasi-parallel strong shocks
\mbox{(e.g., \cite{Caprioli2014,gupta24}),} while the onset of shock-drift acceleration in quasi-perpendicular shocks can act as an effective injection mechanism for electrons (e.g., \cite{guo14}). 
On the other hand, the extrapolation of the basic DSA formalism in Equation (\ref{eq:hb}) has proven to model reasonably well the diffuse and polarized synchrotron emission detected beyond the scale of clusters of galaxies, in a statistical way (e.g., \cite{2021MNRAS.505.4178V,vern23}), and hence we also rely on this to produce forecasts of radio emission elsewhere in this work.

Upper limits of the average magnetic field of the Universe and in cosmic filaments were obtained by cross-correlating all-sky radio maps with cosmic web tracers such as galaxy redshift surveys or hydrodynamics simulations reproducing the nearby cosmic web. \mbox{Ref. \citep[][]{vern17}} used maps at 180 MHz obtained with the Murchison widefield array (MWA) \citep[][]{2013PASA...30....7T} and redshift surveys as tracers, obtaining upper limits in the range 30--2000 nG. Ref. \citep[][]{2017MNRAS.468.4246B} used the map at 2.3 GHz of the project S-PASS \citep[][]{2019MNRAS.489.2330C} observed with the Parkes radio telescope and cosmological simulations as a tracer, obtaining upper limits of 30-130 nG in filaments (the latter value is for a density-weighted cross-correlation), and an upper limit on primordial magnetic fields of 1 nG.

Using LOFAR observations at 144 MHz of two filaments, connecting massive galaxy clusters separated by $\approx10$ Mpc, Ref. \citep[][]{2021A&A...652A..80L} derived an upper limit of the magnetic field in filaments of 250 nG. By stacking LOFAR data at 144 MHz of 106 filaments, an upper limit of 75 nG was found in cosmic filaments \citep[][]{2023MNRAS.523.6320H}. 

The synchrotron emission by filaments was detected at a 5-$\sigma$ by stacking nearly \mbox{390 {,}
000 filaments} connecting galaxy cluster pairs with a projected separation of 1--15 Mpc \citep[][]{2021MNRAS.505.4178V}. Low-frequency data at 73-154 MHz of the experiment LWA \citep[][]{2018AJ....156...32E} and the project GLEAM \citep[][]{2017MNRAS.464.1146H} at MWA were used, along with X-ray data from ROSAT \citep[][]{1992QJRAS..33..165T} in the energy range 0.4--2.4 keV. The clusters were identified using luminous red galaxies from the Sloan Digital Sky Survey (SDSS). They found an average magnetic field in filaments in the range 20--60 nG using equipartition, inverse Compton, and comparison with simulation arguments. The implications for magnetogenesis scenarios, either primordial or astrophysical, will be discussed in Section \ref{sec:discussion}. 
Ref. \citep[][]{2023SciA....9E7233V} repeated a similar analysis using polarized radio maps at 1.4 and 23 GHz with the surveys GMIMS High Band North \citep[][]{2021AJ....162...35W} and WMAP \citep[][]{2013ApJS..208...20B}. Those authors obtained a 3-$\sigma$ detection of the polarized emission from filaments with a large polarization fraction. The latter means that the filament field is highly ordered, or compressed, which supports the current theoretical framework of filaments accreting by accretion shocks of matter infalling from voids.

\subsection{Faraday Rotation Measure}
\label{RM}

The Faraday rotation rotates the polarization angle of a linearly polarized radiation as this travels through a magneto-ionic medium (free electrons in a magnetic field), (e.g., \cite{1966MNRAS.133...67B}). The rotation occurs for the medium turns birefringent and the two circularly polarized components of the radiation propagate at different speeds.
The rotation angle $\Delta \phi$ [rad] is proportional to the wavelength $\lambda$ [m] squared: 
\begin{equation}
\Delta \phi = {\rm RM} \,\lambda^2.
\end{equation}
 {The} 
rotation measure RM is: 
\begin{equation}
{\rm RM} = 0.812 \int_{\rm source}^0 \frac{n_e\,B_\parallel}{(1+z)^2} \,dl\,,
\label{eq:rm}
\end{equation}
where $z$ is the source redshift, the integration is performed from the source to the observer along the path length $l$ [pc] on the sight line, $n_e$ is the free-electron number density [cm$^{-3}$], and $B_\parallel$ is the magnetic field along the line of sight [$\mu$G]. Thus, RM bears information on the magnetized medium the radiation propagates through and is used to estimate the magnetic field properties in several environments, such as in the galaxy \citep[][]{2016A&A...596A.103P, 2022ApJ...940...75D}{}{}, the circumgalactic medium (CGM) of galaxies \citep[][]{2023A&A...670L..23H, 2023A&A...678A..56B}{}{}, the environment local to the source \citep[][]{2008MNRAS.391..521L}{}{}, and the cosmic web \citep[][]{2022MNRAS.515..256P, 2023MNRAS.518.2273C}{}{}. 

It should be noted that a number of complications may make the application of Equation (\ref{eq:rm}) inaccurate. First, the electron number density and magnetic fields do not need to be well correlated at all scales (i.e., below the resolution limit of the observations, or the spatial resolution of simulations used to model them). Second, the application of Equation (\ref{eq:rm}) deserves caution when the magnetic field is not perfectly isotropic, and the distribution of its components is not Gaussian. Discrepancies here are expected in particular environments, such as the multi-phase and dense cores of clusters of galaxies, or else in the presence of large clumping factors in the gas distribution (e.g., \cite{2019MNRAS.484.1427C,2019MNRAS.490.1697O}). Third, in addition to the thermal electrons, non-thermal
electrons can also contribute to the observed amount of Faraday rotation. The likely presence of cosmic rays in large-scale structures, in general, has been highlighted in the latest years (e.g., \cite{pf07,scienzo,scienzo16,wu24,boss24,hopkins25}). { {In} 
clusters of galaxies, a budget above $1\%$ of the thermal gas energy density can probably be excluded based on the lack of detected $\gamma$-ray hadronic emission in clusters of galaxies (e.g., \cite{fermi14,2014ApJ...795L..21G}). On the other hand, on the scale of filaments a larger population of cosmic ray protons might be accommodated, considering that for magnetic fields $\sim10 \rm nG$ (i.e., of the same order of what is implied by our interpretation of radio observations of this work) protons with $E \leq 10^{18} \rm eV$ can be magnetically captured when intercepted by a filament, even if produced elsewhere \mbox{(e.g., \cite{kim19,wu24}).}} The presence of a diffuse distribution of relativistic electrons, injected by shocks and galactic activity, is also very likely, as shown by recent numerical simulations also used in this work (e.g., \cite{va25a}).
We notice that all aforementioned effects can be properly taken care of by using a covariant cosmological polarized
radiative transfer, as suggested by Ref. \cite{2019MNRAS.484.1427C,2019MNRAS.490.1697O}. However, our more conventional approach is suitable in the absence of 
emission and absorption, and for a negligible distribution of non-thermal electrons, all conditions that should be fairly well met in the intergalactic medium we target in our study.

The RM of an extragalactic source is the combination of the RM components along the sight line: 
\begin{equation}
{\rm RM = GRM + RM_{eg}} + N,
\end{equation}
which comprises a galactic RM (GRM) component, an extragalactic term (RM$_{\rm eg}$), and the noise ($N$). The extragalactic term can be produced in the environment local to the \mbox{source \citep[][]{2008MNRAS.391..521L},} in intervening astrophysical objects, such as galaxy clusters or galaxies \citep[][]{2025A&A...693A.208C}, or in the intervening cosmic web \citep[][]{2025A&A...693A.208C}. To investigate the extragalactic component, the GRM has to be subtracted off, giving the residual RM (RRM): 
\begin{equation}
\rm RRM =RM - GRM. 
\end{equation} 

The extragalactic component is dominated by the local term at gigahertz frequencies and by the cosmic web at low frequency (144 MHz) \citep[][]{2022MNRAS.512..945C}, at which the astrophysical component (local term and intervening astrophysical objects) contribution is 21 percent and is only due to intervening objects, while no significant local term is observed \citep[][]{2025A&A...693A.208C}. 
The latter paper also finds that the polarized emission only propagates in low-density environments at low frequency, which can explain why the cosmic web component dominates. 

The RM signal from cosmic filaments was detected by using LOFAR data at 144 MHz, and the average magnetic field strength and its evolution with redshift were obtained by fitting the evolution with the redshift of the LOFAR RRM rms \citep[][]{2022MNRAS.512..945C, 2023MNRAS.518.2273C, 2025A&A...693A.208C}. The fitting draws the electron number density from cosmological simulations, includes an astrophysical component (for more details, see Section \ref{best_fit}), and assumes that the proper magnetic field strength ($B$) evolves with redshift as a power-law: 
\begin{equation}
B=B_0\,(1+z)^\alpha. 
\end{equation}

The best-fit strength at redshift $z=0$ is in the range $B_0\approx10$--40 nG, while the slope is $\alpha \approx 1.2$--2.6, the range mostly depends on the evolution with redshift of the astrophysical component. The comparison with magnetohydrodynamics (MHD) simulations favors an astrophysical component that increases with redshift, which gives $B_0\approx10$--15 nG and $\alpha \approx 2.3$--2.6. This comparison also shows that primordial models are favored compared to astrophysical scenarios, and it is consistent with an initial magnetic field of $\approx 0.37$ nG.

The magnetic field has also been detected in the cosmic web moving away from galaxy groups \citep[][]{2024MNRAS.533.4068A}. It was done using data from the project POSSUM \citep[][Gaensler et al. submitted]{2010AAS...21547013G} at 944 MHz. The RRM profile as a function of the separation to the groups was obtained using background polarized sources of 55 nearby galaxy groups. Besides the RRM signal in the intra-group medium, they estimated fields of 200--600 nG just off the groups, while the signal drops below their sensitivity in the more distant cosmic web. The same signal is not detected at a lower frequency of 144 MHz \citep[][]{2025A&A...693A.208C}, possibly for the impact parameters investigated at 944 MHz are too close to the groups and hence at a too high density, at which the low-frequency signal is depolarized. 

The mean magnetic field of the Universe on very large scales has been investigated using RMs and RM pair differences. The strictest upper limit was obtained using LOFAR data in an RM differential experiment between close pairs of sources, setting an upper limit of 1.5 nG \citep[][]{2022MNRAS.515..256P}. The strictest upper limits on the strength of the primordial magnetic field are set with RRM in filaments, yielding $\leq \rm nG$ (comoving) for large-scale fields. In general, these limits are a function of the parameters of the primordial models, and we refer to \mbox{Ref. \cite{2024arXiv241214825N}} for a further discussion of the possible values and limits.

Table \ref{table:tab1} reports an updated summary of all important constraints of the magnetic field strength outside of clusters of galaxies, obtained with different radio techniques and radio telescopes, and referred to the cosmic overdensity they mostly probed. Each probe reported here is subject to specific biases and contaminants, and each one is sensitive to different properties of the magnetic field (i.e., intensity squared versus component projected along the line of sight). Moreover, a range of different assumptions was used in each work to infer magnetic field properties from observations (e.g., assuming equipartition or more sophisticated emission models, using purely random and single-scale models of magnetic fields, or instead three-dimensional cosmological simulations). 
However, a consistent general trend is that the strength of magnetic fields in the cosmic web on average increases with increasing cosmic overdensity.
This trend is in line with the most basic expectations of theoretical models (see  {next Section)} 
and it can also be used to constrain the possible histories of the magnetization of the Universe.

\begin{table}[H]
\caption{ {Recent} 
relevant measures (or upper limits) of cosmic magnetism outside of clusters of galaxies, based on the radio window alone, sorted by the (approximate) overdensity range they probed. We used the following abbreviations: Sync. = synchrotron emission; CC. = cross-correlation; Pol. = polarized emission; $\Delta RM (\theta)$ = difference in RM as a function of angle in the plane of the sky; RRM = residual rotation measure.}
\footnotesize
\begin{adjustwidth}{-\extralength}{0cm}
	\begin{tabularx}{\fulllength}{cccccC}
	\toprule
\bf{Technique} 
& \bf{Targets} & {\bf Magnetic Field} & {\bf Density Range} & {\bf Instrument(s)} & {\bf References}\\ 
& & \boldmath{$[\rm Comoving~ nG]$} & \boldmath{$\rho/\langle \rho \rangle$} & & \\
\midrule
Sync. & cluster  {bridges}
& $\sim 200-500$& $\sim 50-200$ & LOFAR-HBA (120 MHz) & \cite{2019Sci...364..981G} \\ 
Sync. & cluster pairs & $\leq 250$ & $\sim 5-50$ & LOFAR-HBA (120 MHz) & \cite{locatelli21} \\
Sync. Pol. & cluster pairs & $\geq 0.4 $ & $\sim 300$ & LOFAR-HBA (144 MHz)& \cite{Balboni23}\\
Optical-radio CC. & galaxies & $\leq 250$ &$ \sim 10-10^2 $ & MWA-EoR0 (180 MHz) & \cite{vern17,brown17}\\
RRM & superclusters& $\sim 11-69$ & $ \sim 3-100 $ & VLA(1.4 GHz),LOFAR-HBA(144 MHz) &\cite{2025arXiv250308765P}\\
Sync. stacking & cluster pairs & $\sim 20-60$ & $\sim 5-50 $ & MWA+LWA (50-120 MHz) & \cite{2021MNRAS.505.4178V}\\
Sync. stacking & cluster pairs& $\leq 75$ & $\sim 5-50 $ & LOFAR-HBA (120 MHz)&\cite{2023MNRAS.523.6320H}\\
Pol. stacking & cluster pairs& $\sim 40-60$ & $\sim 5-50 $ & GMIMS (1.4 GHz), PLANCK (30 GHz)&\cite{vern23}\\
$\Delta RM (\theta)$ & radio gal. pairs & $\leq 40$& $\sim 1-10 $ & VLA-NVSS (1.4 GHz) & \cite{vern19} \\
$\Delta RM(\theta)$ & radio gal. pairs &$\leq 4$ & $ \sim 1-10 $ & LOFAR-HBA (120 MHz) & \cite{os20} \\
$\Delta RM(\theta)$ & radio gal. pairs &$\leq 9 $ & $ \sim 1-10 $ & LOFAR-HBA (120 MHz) & \cite{2022MNRAS.515..256P} \\
$RM$ CC. & bg pol. sources &$\leq 30$ & $ \sim 1-10 $ & VLA-NVSS (1.4 GHz) & \cite{2021MNRAS.503.2913A} \\
$RRM(z \leq 3)$& bg pol.sources & $\sim 8-26$ & $\sim 10$ & LOFAR-HBA (144 MHz) & \cite{2023MNRAS.518.2273C}\\
Sync. & full-sky &$\leq 10^{-3}-0.3$ & $ \sim 1 $ & ARCADE2+LW1 ($78 \rm ~MHz$)& \cite{2021EPJC...81..394N} \\ 
\bottomrule
\end{tabularx}
\end{adjustwidth}
\label{table:tab1}
\end{table}

\section{Cosmological Simulations of Large-Scale Magnetic Fields}
\label{sec:sim}

Modern cosmological simulations represent a powerful tool to couple models of the origin and evolution of magnetic fields with the process of hierarchical structure formation. 

Pioneering studies in this respect were conducted by Ref. \cite{1997ApJ...480..481K,1999ApJ...518..594R}, while the first systematic campaign with cosmological simulations was produced by Ref. \cite{do99}. Early simulations typically evolved initial uniform seed magnetic fields (thus mimicking large-scale primordial scenarios) with ideal MHD on a comoving space, which allowed adiabatic changes and amplification of magnetic fields across the cosmic web
(e.g., \citep{do01, br05, 2011MNRAS.418.2234B,ruszkowski11,va14mhd}).
A consistent finding of these studies was that the magnetic field strength must correlate with gas density ($B \propto \rho/\langle\rho \rangle ^{2/3}$) from flux freezing, to a good approximation), except within the small volume of overdense matter halos. 
In self-gravitating matter halos, the
small-scale dynamo amplification breaks the adiabatic compression scaling law, leading first to an exponential growth of magnetic energy, followed by a slower non-linear growth until the magnetic energy reaches an approximate equipartition with the turbulent kinetic energy (e.g., \citep{2004ApJ...612..276S,2014ApJ...797..133C,fed14,2025arXiv250502885B}).
Notwithstanding the technical difficulty of cosmological simulations to simulate a large enough effective Reynolds number, which correctly captures the salient features of the dynamo amplification process \citep[][]{bm16,review_dynamo}, 
the typical field strength obtained through the amplification of small seed fields in simulated clusters is in the $B \sim \mu G$ order of magnitude (e.g., \citep{va18mhd,2020MNRAS.494.2706Q,2022ApJ...933..131S}), in line with radio observations (e.g., \citep{mu04,bo10,bo13,2019SSRv..215...16V}).

\textls[-25]{The apparent success of the ideal MHD simulations here is somewhat surprising, considering that the correct description of a dynamo in a weakly collisional plasma requires sophisticated kinetic physics and is ruled by the onset of instabilities, which cannot be captured by cosmological simulations 
(e.g., \citep{2016PNAS..113.3950R, PhysRevLett.131.055201,zhou2023magnetogenesis, 2018ApJ...863L..25S,2024A&A...683A..35R}). Nevertheless, the view offered by using ideal MHD has provided so far a realistic enough description of the basic properties of magnetism on large scales, at least up to the cluster-wide scales probed thus far by radio observations.}

Unlike the case of the intracluster medium, the plasma within cosmic filaments is not expected to undergo efficient dynamo amplification.
Gas flows in cosmic filaments should be mostly supersonic ($\mathcal{M} \sim 1-10$) and inject compressive turbulence \citep[][]{2016MNRAS.462..448G}. 
Without using full MHD simulations, Ref. \cite[][]{ry08} estimated the maximal magnetization of filaments assuming small-scale dynamo amplification, 
obtaining $B_{\rm rms} \simeq (8 \pi~ \epsilon_{\rm turb} ~\phi)^{1/2}$, where $\epsilon_{\rm turb}$ is the turbulent kinetic energy density, and $\phi$ is a factor that accounts for the growth of magnetic energy under the typical local conditions \citep{ry08}. This assumed that magnetic eddies could be amplified at least for $\sim 10~t_{\rm eddy}$ (where $t_{\rm eddy} \sim l_{\rm eddy}/\sigma_v$ is the eddy turnover time for an eddy with linear size $l_{\rm eddy}$ and a velocity dispersion $\sigma_v$). 
This yields a prediction of $B_{\rm rms} \sim 10-10^2 ~\rm nG$ for the most massive and hot ($\geq 10^7 ~\rm K$) filaments. Based on this, Ref. \cite{2014ApJ...790..123A} predicted a relation for the RM dispersion across different lines of sight crossing a filament's volume:

\begin{equation}
\sigma_{\rm RM} \sim 5 ~\rm rad/m^2 \times \frac{n_e}{10^{-4} cm^{-3}} \times ({\frac{L_{\rm fila}}{5 ~Mpc}})^{1/2} \times ({\frac {l_{\rm eddy}}{0.3 ~Mpc}})^{1/2} \times \frac{B_{\rm rms}}{100 ~\rm nG}
\label{eq:sigma_RM}
\end{equation}
where $L_{\rm fila}$ is the filament thickness.

However, direct MHD simulations have later disputed the above picture as they monitored the small-scale dynamo amplification developed in cosmic filaments and found only modest levels of magnetic field amplification in most of the volume of filaments (e.g., \citep{br05,va14mhd,2015MNRAS.453.3999M,2017CQGra..34w4001V,2018Galax...6..128L}). 
This is motivated by the circumstance that magnetic field lines only have limited amount of dynamical times for their amplification (i.e, less than $\sim 10 t_{\rm eddy}$) because they are advected with 
velocities of several $\sim 10^2 \rm ~km/s$ towards the surrounding halos. Moreover, the available turbulent kinetic energy is predominantly supersonic \mbox{(e.g., \citep{va14mhd,2016MNRAS.462..448G}),} which can only promote low levels of amplification, even in more idealized simulations \citep[][]{fed14}.
Therefore, a large fraction of the volume encompassed by filaments must possess a magnetic field anchored to the input seed field, via compression: 
\begin{equation}
B_{\rm fila,low} \simeq B_{0} \times \left({\frac{\rho} {\langle \rho \rangle}}\right)^{\alpha_B}
\label{eq:Bcomp}
\end{equation}
where $B_0$ is the seed field, $\langle \rho \rangle$ is the cosmic mean (gas) density and $\alpha_B \approx 2/3$ for isotropic gas compression. Cosmic filaments are only mildly non-linear objects of the cosmic web, and their average density is $\rho/\langle \rho \rangle \sim 5-10$ (e.g., \citep{2005MNRAS.359..272C}), so their average compressed magnetic field should be $\sim 3-5$ times larger than the $B_0$ seed field. However, the presence of substructures, supersonic gas motions, and shocks drive gas density fluctuations in this environment, with a distribution of values which can stretch to about 2 orders of magnitude (e.g., \citep{2021A&A...653A.171A}). Therefore, Equation~(\ref{eq:Bcomp}) gives a lower limit on the average magnetic fields in filaments for a given primordial magnetic seed, while their realistic internal distribution of magnetic and density fluctuations can stretch across $1-2$ orders of magnitude.

\begin{figure}[H]
\includegraphics[width=0.32\textwidth]{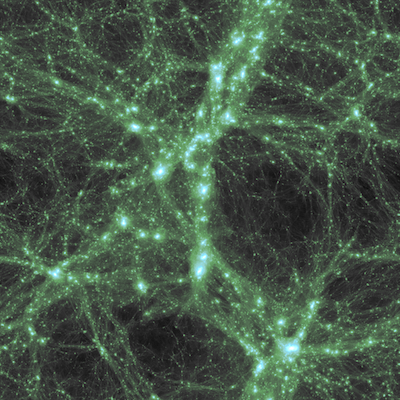}
\includegraphics[width=0.32\textwidth]{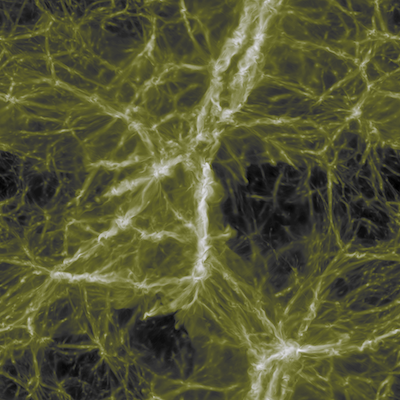}
\includegraphics[width=0.32\textwidth]{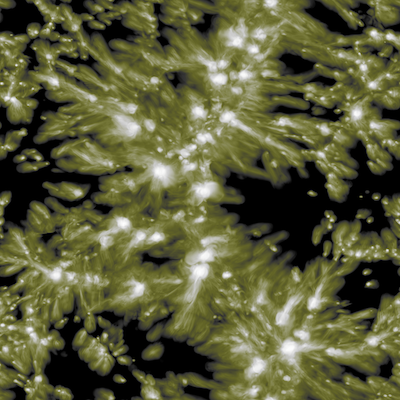}
\caption{ {A} 
visual impression of the different amplitude and filling factors of extragalactic magnetic field strength at $z=0.0$ if the same simulated cosmic volume (with side $85 ~\rm Mpc$) is simulated either starting from a primordial uniform seed field (\textbf{central} panel) or only using astrophysical sources of magnetic seeding (\textbf{right} panel), taken from Ref. \cite{2017CQGra..34w4001V}. The \textbf{left} panel shows the dark matter density projected along the line of sight in the same volume.}
\label{fig:sim_plots}
\end{figure}

Having established that, in filaments (and, even more, in voids), the ``memory'' of the seed fields is likely preserved as their internal dynamics are inefficient to produce dynamo amplification, we turn to considering which seeds of magnetic fields can realistically be there.
A few cosmological simulations have tested the effect of different seeding models of magnetic fields on large-scale structures (e.g., \citep{2009MNRAS.392.1008D,beck12,2015MNRAS.453.3999M,2017CQGra..34w4001V,2019MNRAS.484.2620K,va21magcow}), reporting that even wildly different seeding models can lead to the same magnetic fields in clusters of galaxies at low redshift, while yielding diverging predictions on the larger scales of filaments and voids. An example of the comparison between a primordial seeding model with an astrophysical one (taken from Ref. \cite{2017CQGra..34w4001V}), is given in Figure \ref{fig:sim_plots}: the large-scale distributions of magnetic fields evolved by the cosmological MHD simulation are radically different in the two scenarios, despite the dark matter skeleton underlying both simulations (and hence the gravitational potential) is exactly the same. The volume filling factor of magnetic fields in the primordial model is visibly large, as any pixel in the projected central map would, intercepts some amount of diffuse magnetic field along the line of sight. Conversely, the right map shows that a fraction of pixels in the map (typically far away from matter halos) are black, meaning there are many lines of sight in the cosmic void that do not intercept any magnetic field.

Lately, cosmological MHD simulations simulated the evolution of primordial seed fields with different initial topologies, i.e., by testing the full range of plausible slopes for power-law stochastic seed fields \citep[][]{2021MNRAS.500.5350V}, considering the additional role of primordial helicity \citep[][]{2022ApJ...929..127M} and including seed fields with finite correlation lengths \citep[][]{2024arXiv240616230M}. Along a similar line of research, other simulations attempted to derive the dynamical effects of primordial magnetic fields on the evolution of matter clustering, both on the scales of galaxies \mbox{\citep[][]{2020MNRAS.495.4475M,2024A&A...690A..59S,2024arXiv241002676R},} and on the filaments probed by the Ly-$\alpha$ forest \citep[][]{2025arXiv250106299P}.

\begin{figure}[H]
\includegraphics[width=0.99\textwidth]{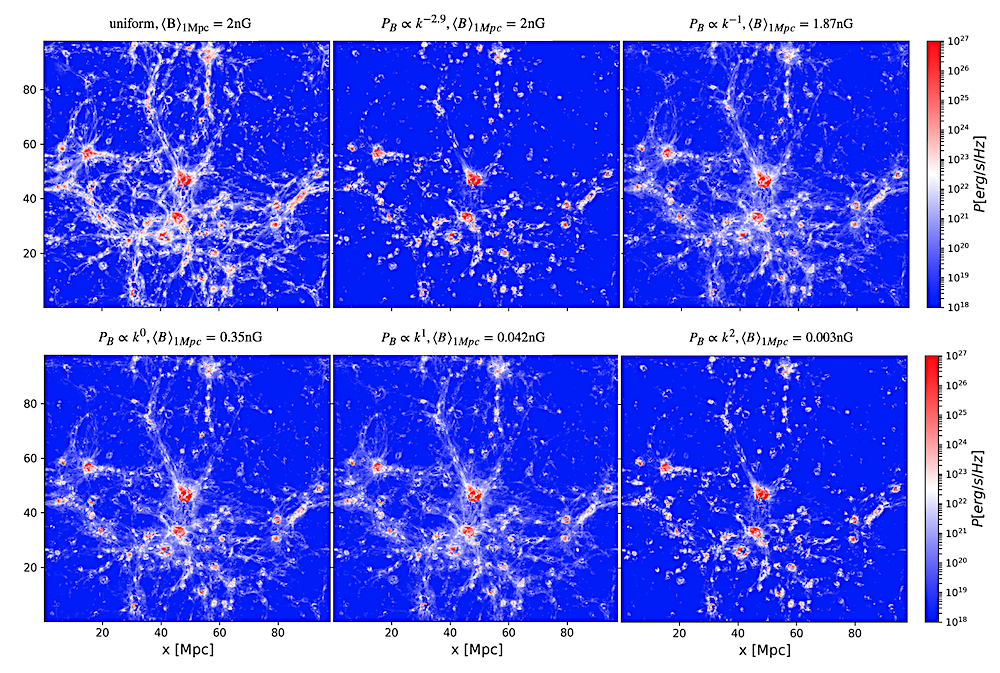} 
\caption{Predicted synchrotron radio power at $100 ~\rm MHz$ emitted by shock-accelerated relativistic electrons in the cosmic web, for resimulations of the same $100^3 \rm Mpc^3$ volume (here at $z=0$), starting from six different models of primordial magnetic fields compatible with CMB constraints, as detailed in \citet{2021MNRAS.500.5350V}.}
\label{fig:sync_all}
\end{figure}

\begin{figure}[H]
\includegraphics[width=0.99\textwidth]{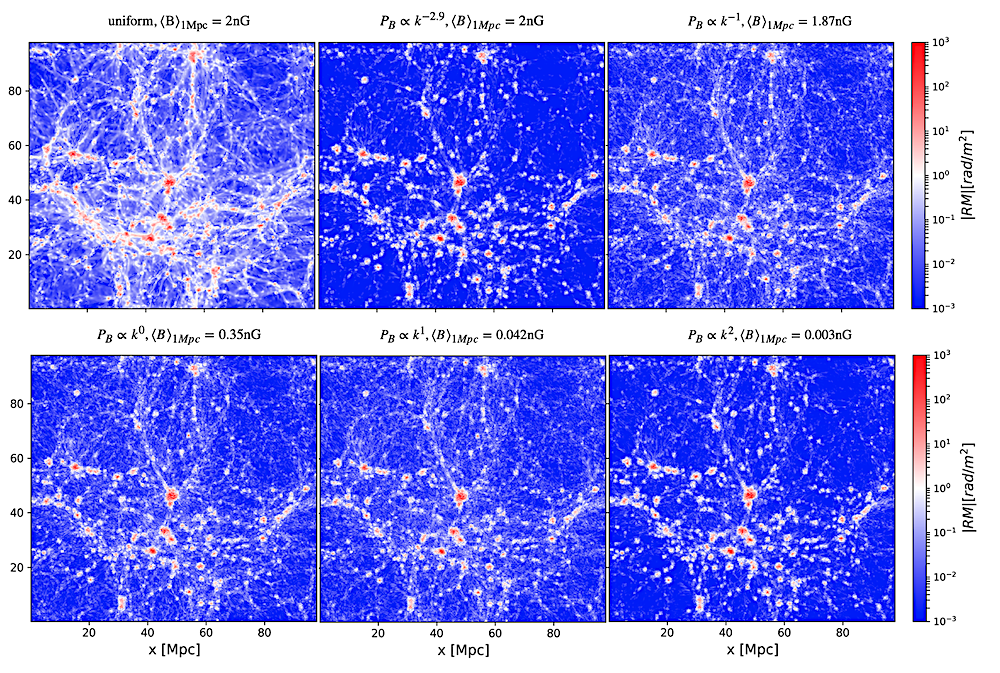}
\caption{Predicted Faraday Rotation at $z=0$ (in absolute value and only integrated for $100 \rm ~Mpc$ along the line of sight) for the same simulated volume of Figure 2. 
}
\label{fig:RM_all}
\end{figure}

\textls[-15]{The fact that primordial models calibrated from CMB constraints (e.g., \citep{2019JCAP...11..028P}) yield different large-scale distributions of synchrotron radio power from cosmic shocks, as well as of integrated Faraday rotation along the line of sight, is the key that makes radio observations a powerful probe of primordial magnetism even using observations of the low-redshift Universe, potentially even more powerful than CMB observations, at least for a range of models. While this will be explored in more detail in Section \ref{sec:analysis}, we show here the status of previous simulations, which predicted the level of radio emission (Figure \ref{fig:sync_all}) and Faraday rotation (Figure \ref{fig:RM_all}) for a large cosmic volume resimulated for six different variations of magnetic field initial spectra, including inflationary-like models, purely causal ones, or an initially homogenous magnetic field. The figures clearly show how, outside of dense halos of the cosmic web, the different combinations of spectral index and normalization allowed by CMB limits yield significantly different distributions of both observables, in filaments and in voids of the cosmic web. In synchrotron, the differences are more prominent and potentially detectable in massive filaments in rich environments (e.g., around or in between massive halos), owing to the rapid drop of the radio power with the shock kinetic energy flux and $B^2$ (Figure \ref{fig:sync_all}). In the Faraday rotation integrated along the line of sight (Figure \ref{fig:RM_all}), the differences are more marked when the entire cosmic volume is considered, which further motivates a statistical analysis of real observations, for large integrated cosmic volumes.}

\textls[-15]{The present-day magnetic fields in cosmic voids are crucial complementary information to further constraints, the origin of cosmic magnetism, on scales that cannot be probed by radio observations. Key here are the non-detections of the inverse Compton cascade (ICC) signal at $1-100 \rm GeV$ for blazars that are instead detected at $\sim \rm TeV$: since the seminal work by Ref. \cite{2010Sci...328...73N}, the consensus interpretation is that the lack of the ICC is due to the deflection of electrons and positron pairs by intergalactic magnetic fields, extended on $\sim 10-100 \rm ~Mpc$. More recently, several modelings have concluded that, regardless on the specific details of implemented recipes for galaxy formation, evolution, and feedback, purely astrophysical scenarios for the origin of cosmic magnetism are 
incapable to pollute a large enough fraction of the volume of voids (i.e., $\geq 60 \%$), to explain the lack of ICC (e.g., \cite{2022MNRAS.515.5673A,2024ApJ...963..135T}).}

On the other hand, cosmological simulations still do not agree on the quantitative expectation for the magnetization of filaments by galaxy feedback, which is crucial to interpreting radio observations. The problem here is partly due to the fact that 
very different numerical models of galaxy evolution are capable to account for many observable properties of galaxies in the local Universe (e.g., \cite{2025arXiv250206954V} (for a recent review)), yet their subgrid models produce very different predictions for the impact of galaxy feedback on the intergalactic medium (e.g., \cite{2025ApJ...980...61M}). As a consequence, the predictions of the magnetic pollution by galaxies onto the surrounding filaments remain poorly constrained.

\subsection*{{Latest} 
Simulations of Cosmic Magnetism}
\label{newsim}
\textls[-15]{In order to better constrain the magnetization of the intergalactic medium by galaxies, we have recently presented in Ref. \cite[][]{va25a} new cosmological simulations with the ENZO code,\endnote{\url{https://enzo-project.org} (accessed on 15/05/2025)} 
calibrated to reproduce several key observables of galaxies and at the same time give better limits on the maximal magnetization by feedback processes (by AGN and star formation). We have already compared these simulations to LOFAR observations in \mbox{Ref. \cite[][]{2025A&A...693A.208C},} showing that no astrophysical magnetization mechanism can, alone, account for the observed RRM rms for $z \geq 1$. Here, we give a short summary of the main details of these recent simulations, considering that we will use them in the remainder of the paper to compare both with RRM rms data (Sections \ref{best_fit} and \ref{askap}) and synchrotron (Section \ref{stacking}) observations.}

The new simulations covered a $42^3 \rm Mpc^3$ comoving volume at the constant spatial resolution of $41.5$ kpc/cell. The galaxy formation routines (i.e., star formation and feedback, feedback from active galactic nuclei) were tailored to reproduce the cosmic star formation history, the distribution function of stellar mass in galaxies, the stellar mass fraction in galaxies, the luminosity distribution of radio galaxies, and the correlation between supermassive black hole mass and the host halo gas mass, as discussed in detail in Ref. \cite{va25a}. 
In ``astrophysical'' seeding models, the feedback by star formation and by AGN was assumed to release magnetic fields corresponding to a fixed ($10\%$) fraction of their energy. In "primordial" models instead, magnetic fields were additionally initialized at the start of the simulation by randomly generating (in Fourier space)
stochastic magnetic fields. The magnetic vectors were drawn a random realization of a power-law spectrum: $P_B(k) = P_{B0}k^{nB}$, while the normalization was set by requiring that the field strength, after smoothing over a fixed scale (e.g., $\lambda=1 \rm ~Mpc$) is consistent with limits from the CMB analysis (e.g., \cite[][]{2021MNRAS.500.5350V}). 
We consider here the $nB=-1$, $=0$, $=1$, and $=2$ cases. 
The normalizations of each seed field were based on Ref. \cite{2019JCAP...11..028P}: $\langle B\rangle _{\rm 1Mpc}=0.003\,\rm nG$ for $nB=2$, $\langle B\rangle_{\rm 1Mpc}=0.042 \, \rm nG$ for $nB=1$, $\langle B\rangle_{\rm 1Mpc}=0.35 \,\rm nG$ for $nB=0$ and $\langle B\rangle_{\rm 1Mpc}=0.37 \, \rm nG$ for $nB=-1$. We notice that the last case is the only one where, based on our previous work \cite{2025A&A...693A.208C}, the normalization is not set by the CMB analysis, but from the comparison with LOFAR RRMs, which yields a normalization $\sim 5$ times below CMB constraints. For completeness, we also consider here a simpler uniform magnetic field model, with initial $B_0=0.1$ nG and $B_0=0.6$ nG (in Section \ref{best_fit} and \ref{stacking}, respectively) imposed to all magnetic field components at the start of the simulation. The latter normalization is because it gave the best results in Ref. \citep[][]{2021MNRAS.505.4178V}.

\section{Analysis}
\label{sec:analysis}
\subsection{Best Fitting of B Properties in Filaments and Expected RRM versus z from Simulations}
\label{best_fit}
We reproduce the analysis conducted in Ref. \citep[][]{2025A&A...693A.208C}, where the average magnetic field ($B$) properties in cosmic filaments are derived by best fitting an analytical relation of $B$ versus the redshift ($z$) to the observed trend of RRM rms as a function of $z$. The RRM rms predicted by a number of magnetogenesis scenarios is also computed using MHD cosmological simulations. The major difference, which motivates this part of the work, is that, here, we use the most updated astrophysical scenario developed in Ref. \citep[][]{va25a}, named B4 there, that was not available at the time of our previous work, and its combination with a primordial stochastic scenario with slope $nB =-1$. The new astrophysical model produces a better star formation history that better fits the observed data at low redshift. 

The best-fitting procedure is fully described in Ref. \citep[][]{2025A&A...693A.208C}. Here, we report the most important features. The equation to fit is: 
\begin{eqnarray}
\left<RRM^2\right>^{1/2} &=& \frac{A_{rrm}}{(1+z)^\gamma} + \left<RRM_f^2\right>^{1/2} ,\label{eq:fitting}\\
RRM_f &=& 0.812\, \int_z^0 \frac{n_e\,B_\parallel}{(1+z)^2}\, dl \, \label{eq:rrmf}
\end{eqnarray}
where $n_e$ [cm$^{-3}$] is the electron number density, $B_\parallel$ [$\mu$G] is the magnetic field parallel to the LOS, and $dl$ [pc] is the differential path length. All units are physical (proper) units. The term RRM$_f$ is the cosmic filament component. The term $A_{rrm}/(1+z)^\gamma$ accounts for an astrophysical component, either local or by intervening objects. 
Depending on the value of $\gamma$, the astrophysical component has a different shape, either increasing ($\gamma < 2$) or decreasing with the redshift ($\gamma > 2$). The filament component is computed, and the rms performed over all the LOS. 

We do not have $n_e$ along the observed LOS, hence this is drawn from the cosmological simulations described in Section \ref{sec:sim}, based on the 100 LOS that are randomly extracted from the realization of each scenario. The LOS at 144 MHz only tends to pass through low-density environments, and it avoids high-density structures, such as galaxy clusters, for high-density regions, depolarizing the low-frequency radiation that travels through them. To mimic this, the high-density points of each LOS are flagged, following the procedure described in Ref. \citep[][]{2025A&A...693A.208C}.

The proper magnetic field strength of the filament component is assumed to follow a power-law as a function of redshift: 
\begin{equation}
B_f = B_{f,0} \, (1+z)^\alpha ,
\end{equation}
where $B_{f,0}$ is the strength at $z=0$. 
The comoving field follows a power-law as well:
\begin{equation}
cB_f = B_{f,0} \, (1+z)^\beta ,
\end{equation}
where $\beta$ is related to $\alpha$ by $\beta =\alpha-2$. The field orientation is randomly set each time the LOS enters a new filament. 

Equation (\ref{eq:fitting}) is fit to the RRM rms as a function of the redshift obtained in \mbox{Ref. \citep[][]{2025A&A...693A.208C}} using the LOFAR Two-Meter Sky Survey (LoTSS) Data Release 2 RM catalog at \mbox{144 MHz \citep[][]{2023MNRAS.519.5723O}.} We use their instance obtained using 20-source bins, and only keeping sources with $\rm GRM <14$ rad m$^{-2}$, to minimize residual GRM contamination. 

\textls[-15]{We performed a least-square best fit using two different shapes of the astrophysical component, $\gamma =1$ and $\gamma =5$. The former gives an astrophysical term increasing with the redshift and was the best match with the RRMs predicted by simulations in our previous work. The latter gives a decreasing term and a match with the observed fractional contribution of the astrophysical component of $21\pm 4$ percent better than the values obtained with the shapes used in our previous work, which are all larger than the observed value.}

\textls[-25]{The best-fitting results for the two astrophysical term shapes are reported in Tables \ref{tab:fit_gamma1} and \ref{tab:fit_gamma5}. For the astrophysical component increasing with the redshift of shape $A_{rrm}/(1+z)$ ($\gamma =1$), the average magnetic field strength in filaments at $z=0$ is in the range $B_{f,0}=11$--14 nG, depending on the magnetogenesis scenario, and a typical statistical error of 4 nG. The slope of the proper field is $\alpha =2.4$--$2.6\pm 0.5$ and a comoving field slope of $\beta =0.4$--$0.6\pm 0.5$, which is consistent with a comoving field strength invariant with redshift. The normalization of the astrophysical contribution is $A_{rrm}=1.01$--$1.07\pm 0.09$ rad m$^{-2}$. 
If we assign to each source a contribution according to its redshift, the RRM rms associated with the term $A_{rrm}/(1+z)$ is \mbox{0.70--$0.75\pm 0.06$ rad m$^{-2}$.} Compared to the RRM rms of the sample of $1.54\pm0.05$ rad m$^{-2}$ \citep[][]{2025A&A...693A.208C}, the astrophysical contribution is 46--$48\pm 4$ percent, which is higher than the observed value.} 

\begin{table}[H] 
\caption{Best-fitting results of Equation (\ref{eq:fitting}) to the RRM rms obtained from LOFAR data at 144 MHz in case the astrophysical component contribution is assumed to be $A_{rrm}/(1+z)$. The columns are the magnetogenesis model and the fitting results: the slope of the behavior of the proper magnetic field strength versus $z$ ($\alpha$)
; the magnetic field strength at $z=0$ ($B_{f,0}$); the normalization of the astrophysical term ($A_{rrm}$); and the slope of the comoving field versus $z$ ($\beta$). The models labeled with $nB$ are primordial stochastic models with the power spectrum of slope $nB$. B4 is the label of the astrophysical model used here in the original work where it was developed \citep[][]{va25a}.\label{tab:fit_gamma1}}

\begin{tabularx}{\textwidth}{lCCCC}
\toprule
\textbf{Model} & \boldmath{$\alpha $} & \boldmath{$B_{f,0}$} & \boldmath{$A_{rrm}$} & \boldmath{$\beta$} \\ 
& & \textbf{[nG]} & \boldmath{\textbf{[rad m}$^{-2}$\textbf{]}} & \\
\midrule
$nB$ =  {$-$}
1 & $2.6 \pm 0.5$ & $11 \pm 4$ & $1.07 \pm 0.08$ & $0.6 \pm 0.5$ \\ 
$nB$ = 0 & $2.4 \pm 0.5$ & $13 \pm 4$ & $1.06 \pm 0.09$ & $0.4 \pm 0.5$ \\ 
$nB$ = 1 & $2.5 \pm 0.5$ & $13 \pm 4$ & $1.06 \pm 0.09$ & $0.5 \pm 0.5$ \\ 
$nB$ = 2 & $2.5 \pm 0.5$ & $13 \pm 4$ & $1.07 \pm 0.09$ & $0.5 \pm 0.5$ \\ 
uniform & $2.5 \pm 0.5$ & $13 \pm 4$ & $1.07 \pm 0.08$ & $0.5 \pm 0.5$ \\ 
astro B4 & $2.4 \pm 0.5$ & $14 \pm 4$ & $1.05 \pm 0.09$ & $0.4 \pm 0.5$ \\ 
astro B4 + $nB$ =  {$-$}
1 & $2.4 \pm 0.5$ & $13 \pm 4$ & $1.01 \pm 0.10$ & $0.4 \pm 0.5$ \\
\bottomrule
\end{tabularx}
\end{table}
\vspace{-12pt}

\begin{table}[H] 
\caption{As for Table \ref{tab:fit_gamma1}, except $A_{rrm}/(1+z)^5$, which is assumed as a shape for the astrophysical \linebreak component contribution.\label{tab:fit_gamma5}}
\begin{tabularx}{\textwidth}{lCCCC}
\toprule
\textbf{Model} & \boldmath{$\alpha $} & \boldmath{$B_{f,0}$} & \boldmath{$A_{rrm}$} & \boldmath{$\beta$} \\ 
& & \textbf{[nG]} & \boldmath{\textbf{[rad m}$^{-2}$\textbf{]}} & \\
\midrule
$nB$ = $-$1 & $0.1 \pm 1.2$ & $54 \pm 18$ & $1.16 \pm 0.24$ & $-1.9 \pm 1.2$ \\ 
$nB$ = 0 & $0.2 \pm 0.7$ & $60 \pm 12$ & $1.14 \pm 0.18$ & $-1.8 \pm 0.7$ \\ 
$nB$ = 1 & $0.3 \pm 0.9$ & $56 \pm 15$ & $1.13 \pm 0.21$ & $-1.7 \pm 0.9$ \\ 
$nB$ = 2 & $0.3 \pm 0.8$ & $57 \pm 13$ & $1.14 \pm 0.20$ & $-1.7 \pm 0.8$ \\ 
uniform & $0.3 \pm 0.8$ & $57 \pm 13$ & $1.16 \pm 0.19$ & $-1.7 \pm 0.8$ \\ 
astro B4 & $0.5 \pm 0.6$ & $54 \pm 10$ & $1.17 \pm 0.17$ & $-1.5 \pm 0.6$ \\ 
astro B4 + $nB$ = $-$1 & $0.8 \pm 0.5$ & $43 \pm 7$ & $1.08 \pm 0.17$ & $-1.2 \pm 0.5$ \\ 
\bottomrule
\end{tabularx}
\end{table}

For the astrophysical component decreasing with the redshift of shape $A_{rrm}/(1+z)^5$ ($\gamma =5$), the results are $B_{f,0}=43$--$60\pm13$ nG, $\alpha =0.1$--$0.8\pm 0.8$, $\beta =[-1.9$, $-1.2]\pm 0.8$, and $A_{rrm}=1.08$--$1.17\pm 0.20$ rad m$^{-2}$. The field is stronger and the slope flatter than in the previous case, and it is consistent with a proper magnetic field invariant with the redshift. The fractional contribution of the astrophysical component is 21--$23\pm 4$ percent, which is consistent with the observed value of $21\pm4$ percent.

\begin{figure}[H]
\includegraphics[width=0.8\textwidth]{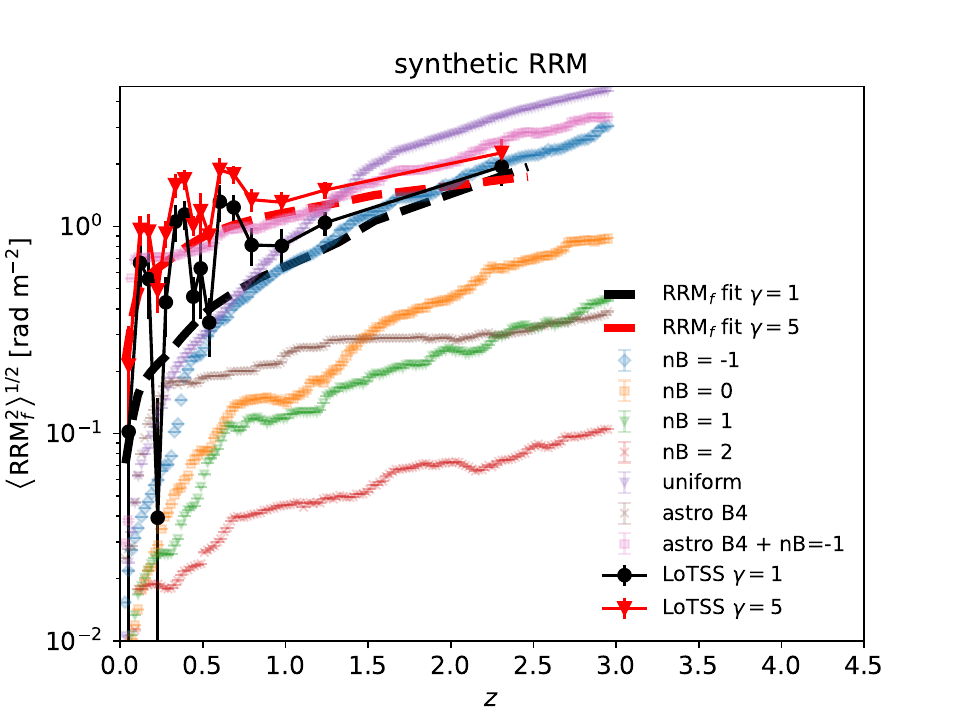}
\caption{ {RRM} 
rms as a function of the redshift obtained with our MHD simulations of magnetogenesis scenarios. The labels are as for Table \ref{tab:fit_gamma1}. The observed values obtained with LOFAR (LoTSS) data are also shown with the astrophysical component subtracted off for two shapes, $A_{rrm}/(1+z)$ (black, solid line) and $A_{rrm}/(1+z)^5$ (red, solid line). The best fits of the filament component (Equation (\ref{eq:rrmf}))) are also reported (black and red, dashed lines for $\gamma =1$ and 5, respectively).}
\label{fig:rmsynth}
\end{figure}

We also computed RRM rms as a function of the redshift for each magnetogenesis scenario in Section \ref{newsim} using the electron number density and the magnetic field from the simulations, to be compared with the observed trend at low frequency. We used the \mbox{100 LOS} extracted from each simulation and computed RRM rms with a redshift resolution of $\Delta z=0.02$, to better inspect the low redshift end. The high-density pixels are flagged with the same procedure used for the best fitting and for the same reason. 
The results are shown in Figure \ref{fig:rmsynth}. We also report the observed RRM rms with the astrophysical component contribution $A_{rrm}/(1+z)^\gamma$ subtracted off for the two shapes that we study in this work, $\gamma = 1$ and $\gamma =5$. The value of $A_{rrm}$ is set to the highest value of Table \ref{tab:fit_gamma1} or \ref{tab:fit_gamma5}, depending on the shape. 

All simulated models show a gentle drop towards $z=0$, except the primordial stochastic model combined with the new and most sophisticated astrophysical model (B4 + $nB=-1$) that stays high down to a very low redshift, and then it quickly drops. 
The reasons for this behavior will be better explained in Section \ref{askap}: feedback at low redshift, both from active galactic nuclei and star formation, seeds new magnetic field in the cosmic web, but it also transports to a larger radius the primordial magnetic field which were previously compressed and amplified in the halos of the cosmic web. This makes it more likely for a simulated LOS to intercept the peripheral regions of bubbles around galaxies, which contain both primordial and astrophysical magnetic fields. Overall, including primordial magnetic fields, it contributes to a $\sim 5-10$ times higher RRM, compared to the most realistic astrophysical model in our suite of simulations.

\subsection{Synchrotron Emission Stacking in Filaments}
\label{stacking}


Using the latest runs of Section \ref{newsim}, we also produced a new important consistency check by testing whether the models that best explain the RRM pattern observed by \mbox{Ref. \cite{2025A&A...693A.208C}} can also account for the existing detection of stacked synchrotron emission \cite{2021MNRAS.505.4178V,vern23}. 

The simulated RRM is taken from the previous Section. Since the RRM at the highest available redshift yields the largest difference between tested models, for simplicity, we shall focus in this new analysis on the final RRM values at $z=2.5$, in all models, and compare with the same values obtained in Ref. \cite{2025A&A...693A.208C} using LOFAR. 

For the radio stacking test, we simulated synchrotron emission generated by relativistic electrons accelerated by shocks, as in Ref. \cite{va25a} and based on Equation (\ref{eq:hb}). We selected a sample of $\sim 30$ filaments connecting the most massive halos in the simulation, separated by $\leq 10 \rm ~Mpc$ in projection. We stacked their emission ($F_{\rm stack,118MHz}$) at $118 ~\rm MHz$, considering a projected area similar to the resolution beam used by Ref. \cite{2021MNRAS.505.4178V}. For simplicity, we assumed that all filaments are at the average cosmological redshift of the objects in the catalog used by Ref. \cite{2021MNRAS.505.4178V}, i.e., $z =0.14$. 

In the following, we will jointly test the simulated RRMs at $z=2.5$ and the filament stacking at $z=0.14$, $F_{\rm stack,118MHz}$, against the real observations. 
Clearly, these two observables provide a very different temporal and spatial sampling of magnetic fields in the Universe. Moreover, they are produced by different parts of the distribution of cosmic filaments: RRM probes the integrated distribution of filaments over cosmologically long lines of sight ($\sim \rm Gpc$) and the signal is dominated by short, low-mass and cold filaments ($T \leq 10^5 \rm K$ , e.g., (e.g., \cite{gh15}). The stacked synchrotron emission instead targets, by selection, massive and hot ($\sim 10^6-10^7 \rm K$) filaments between massive pairs of halos at low redshift. 
This can actually be regarded as a strength of this joint test: by checking for the signature of cosmic magnetism on very different scales, redshift, and densities, we can constrain magnetic field scenarios in a more stringent way, i.e., without focusing on a specific (and potentially biased) set of objects. 

\begin{figure}[H]
\includegraphics[width=0.8\textwidth]{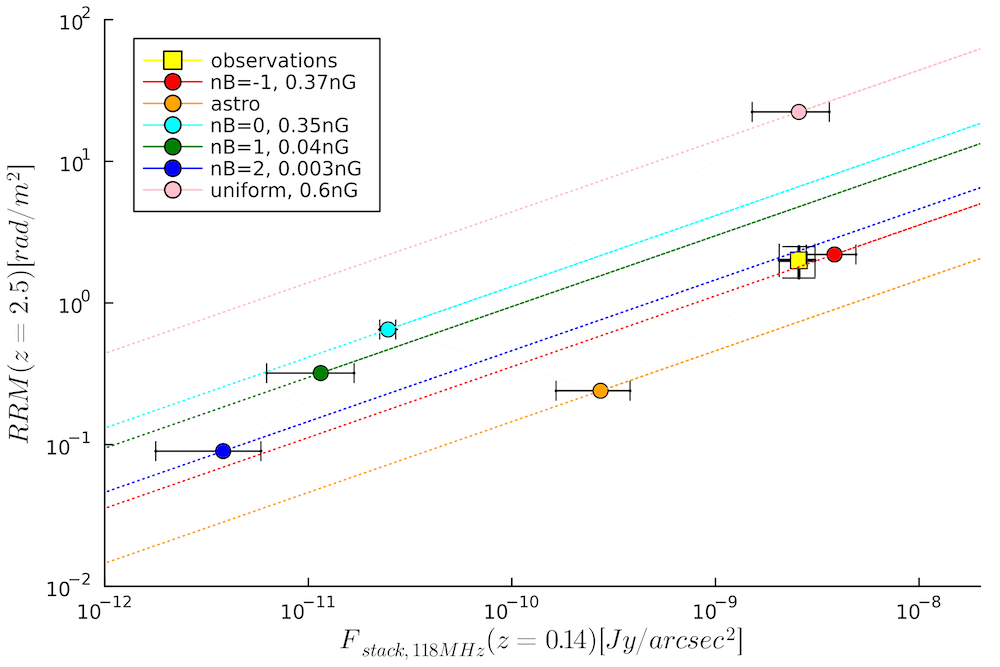}
\caption{
 {Residual} 
RM at $z=2.5$ versus synchrotron radio surface brightness of sample of filaments stacked at $z=0.14$, for five primordial models and for one astrophysical model (colored circles), compared with the result of observations (yellow square, based on  {Ref.} 
\cite{2025A&A...693A.208C,2021MNRAS.505.4178V}) The dotted lines show the approximate effect of renormalizing the seed magnetic fields in all simulations, regardless of existing CMB limits.}
\label{fig:new_sim}
\end{figure}

Figure \ref{fig:new_sim} shows the result of this test. 
Clearly, one of our models stands out, being the most reasonably close to the observations: the stochastic primordial model with $nB=-1$ and $\langle B \rangle_{\rm 1Mpc}=0.37 \rm ~nG$. All other tested models fail to match one or both observables.

We reiterate that the stochastic $nB=-1$ model (which connects to a primordial inflationary generation) has an initial amplitude $\sim 5$ times below the CMB upper limits, and is also compatible with lower limits from the non-detection of ICC from blazars. This model implies very large correlation scales. In the low magnetization level necessary to explain both observables in the cosmic web ($\ll \mu \rm G$), the emitted synchrotron power is degenerate with respect to $B^2$ and $\xi_e$, i.e., the acceleration efficiency of relativistic electrons, which is a fraction of the kinetic energy flux across shocks. This is typically $\sim 10^{-2}$ in our model, corresponding to a fraction of accelerated electrons out of the thermal pool of order $\sim 10^{-4}$ of the number of thermal electrons. These numbers are in line with our best current knowledge of diffusive shock acceleration in space plasmas (e.g., \cite{gupta24}), and it is thus unrealistic that the real efficiency can be much larger than this. 
If we neglect the existing priors from CMB, we can just 
rescale in post-processing the amplitude of the initial magnetic field in all tested models, and assess the scaling factor, which can allow each model to approximately match observations. 
The effect of this rescaling is shown by the oblique line-crossing model: they show how the stacked emission (which approximately scales with $\propto B^2$) and the RRM ($\propto |B|$) are changed by the rescaling. Interestingly, except for the causal $nB=2$ model, no arbitrary renormalization of the seed fields can reproduce RRM and the synchrotron stacking at the same time (besides the $nB=-1$ model). 
Only for $nB=2$, a seed field $\sim 10-15$ times higher than what is allowed by CMB limits (i.e., $\sim 0.03-0.045 \rm ~nG$ instead of $\sim 0.003 \rm ~nG$) might in principle produce RRM and a stacked radio surface brightness in line with observations. Based on Ref. \cite{2025A&A...693A.208C}, we can also conclude that a rescaled $nB=2$ model might also approximately reproduce the full trend of $RRM(z)$ from $z=0$ to $z=2.5$, besides the $z=2.5$ datapoint considered here. 

Unlike the $nB=-1$ model, in the $nB=2$ model, the magnetic energy peaks on the smallest scale, thus implying a causal generation of magnetic field, as 
in post-inflationary models (i.e., Electro-Weak or QCD transitionary models, e.g., \cite{2016PhyS...91j4008K,2016RPPh...79g6901S,2021RPPh...84g4901V}).
Finite numerical methods can artificially suppress magnetic energy on scales close to the resolution limit, which can be a problem for this model in particular: based on Ref. \cite[][]{2019JCAP...11..028P}, the initial magnetic field in the $nB=2$ model has structure down to $\sim 10 ~\rm kpc$ scales, while the resolution of our simulations is $41.5 ~\rm kpc/cell$. Therefore, a fraction of the initial energy spectrum is truncated since the start, and in addition, a fraction of the magnetic energy on scales slightly above the resolution limit might have been artificially dampened by the simulation. This might artificially lower both the RRM and the stacked emission, even if it seems implausible that finite resolution effects, alone, can account for the required $\sim 100-200$ difference in magnetic energy to match the observations. 

We will further discuss the implications of this result in Section \ref{sec:discussion}.

\subsection{Rotation Measure Profiles in Galaxy Groups} 
\label{askap} 
\begin{figure}[H]
\includegraphics[width=0.71\textwidth]{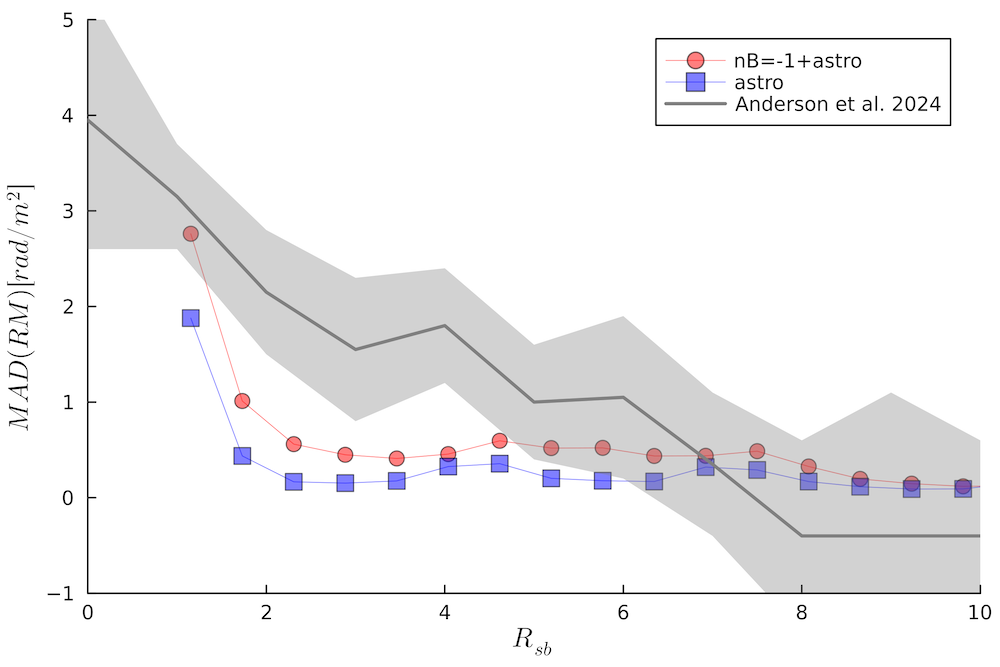}
\caption{
Radial profile of the mean absolute deviation for RRM values around $\geq 3 \times 10^{12} M_{\odot}$ groups of galaxies in two simulated magnetogenesis scenarios (the preferred astrophysical model of Ref. \cite{va25a} with or without and addition primordial fields with a $nB=-1$ spectrum, as in Figure \ref{fig:new_sim}), compared to the data from the ASKAP observation by Ref. \cite{2024MNRAS.533.4068A} (gray lines shaded area). In the latter dataset, we removed the contribution local to the extragalactic background sources of $6.4 \rm ~rad/m^2$ estimated by those same authors.}
\label{fig:new_MAD}
\end{figure}

Recently, Ref. \cite{2024MNRAS.533.4068A} used POSSUM RRMs observed at 944 MHz with the telescope ASKAP to obtain the radial profile of the median absolute deviation (MAD) of RRMs around groups of galaxies in the nearby Universe ($z \leq 0.1$). We replicated this measurement using the same simulations of the previous sections, in which we extracted a total of \mbox{$\sim 70$ halos} (using three different snapshots of each run for $z \leq 0.1$) in the $M_{\rm 100} \geq 3 \times 10^{12} M_{\odot}$ mass range, to obtain a sample similar to Ref. \cite{2024MNRAS.533.4068A}. We focus here on the comparison between the best astrophysical model of \citet{va25a}, with or without the addition of the primordial field with a $nB=-1$ spectrum considered above. 
To compare simulations and observations in a consistent way, we quadratically subtracted the estimated local contribution of extragalactic RRM of $6.4~ \rm rad$ m$^{-2}$ (as estimated by those authors) from the POSSUM MAD measurements. Following the approach of the original work by Ref. \cite{2024MNRAS.533.4068A}, we normalized all radii of simulated groups to the ``splash back radius'' ($R_{sb}$), computed based on the mass of each halo, which is a convenient parametrization of the radius of groups and clusters of galaxies in the cosmological context (e.g., \cite{2017ApJ...843..140D}).
Our results are shown in Figure \ref{fig:new_MAD}.

\textls[-15]{Outside $\sim R_{sb}$ both simulated models fall short of the trend measured with ASKAP, approximately by a factor $\sim 2-3$ in the primordial model and $\sim 3-6$ in the case of the sole astrophysical scenario. Starting from about $\sim 5 R_{sb}$, the MAD RRM in the primordial model is compatible within the errors, while for $\geq 7 R_{sb}$ both are close to 0. This test shows that the combination of primordial and astrophysical models, which best fits RRM and the stacking of the radio surface brightness explored above, remains the closest to the recent ASKAP observations of galaxy groups, while even our most efficient sole astrophysical scenario falls short here. On the other hand, some physical mechanisms, or numerical details, must be missing in the simulated model. The mismatch of a factor $\sim 2$ necessary to reproduce the observed MAD RRM can plausibly
be obtained in several ways, e.g., with a clumpier medium in higher resolution simulations, by an increased number of feedback sources around groups than what is captured by this simulation by a slightly higher feedback duty cycle, or by considering once more the effects of the limited spatial resolution of our model, which is likely to underestimate the amount of small-scale amplification in these overdense regions.}

\section{Discussion}
\label{sec:discussion}

We performed a comparison between radio observations of large-scale 
emission and RRM with the predictions of recent cosmological simulations of a number of magnetogenesis models, with the goal of testing which one of the major scenarios (i.e., astrophysical or primordial) is favored by the data. 
Our tests focus on cosmic filaments, which are the most rarefied structures of the cosmic web where magnetic fields have been recently detected, with very low ($\sim \rm nG$) strength. The even more rarefied regions of voids remain out of reach and require complementary techniques (e.g., high-energy blazar observations). 

The analysis of the RRM rms as a function of the redshift shows that the primordial models, generating fields from the early stages of the Universe, can be large enough at high redshift to match the slope of the observed data, as found in Ref. \citep[][]{2025A&A...693A.208C}. This also occurs for the combined model of the primordial stochastic model with slope $nB=-1$ and the new astrophysical model explored here. The normalization of models with slopes $nB \geq 0$, already set to the current upper limits from CMB observations \citep[][]{2019JCAP...11..028P}, provides instead a RRM rms significantly smaller than that observed. The amplitude of RRM rms of the scenario $nB=-1$ well matches the observed RRMs, provided that the normalization is set $\approx 5$ times lower than the existing upper limit from CMB observations. The $nB=-1$ model is therefore the only scenario that is consistent with both the best constraints of cosmic magnetism from the $z \leq 3$ cosmic web and the CMB limits coming from $z \sim 1000$. This result is particularly intriguing, as it shows that the advanced modeling of RRM rms values in the low-density Universe has become more sensitive than CMB analysis to probe primordial magnetic fields for inflationary spectra, as also discussed in Ref. \cite{2024arXiv241214825N}. 

At low redshift, there is a good match between data and simulations for both astrophysical and primordial scenarios if $A_{rrm}/(1+z)$ ($\gamma =1$) is assumed as the shape of the astrophysical component. However, our preferred model is the combined scenario: it matches the observed RRM rms at high redshift and also includes the local magnetization bubbles resulting from the galaxy feedback predicted by the astrophysical model, which must be present. In such a case, there is a good match between observations and simulations if $A_{rrm}/(1+z)^5$ ($\gamma =5$) is assumed as the shape of the astrophysical component. 
This shape also has the relevant benefit of predicting a fractional contribution of the astrophysical component that is consistent with the observed $21$ percent. 

Considering the match between models and observations, based on the analysis of the sole RRM rms, we can conclude that there is: 
\begin{itemize}
\item A primordial magnetic field, while the astrophysical model cannot match the observations at high redshift; 
\item A filament magnetic field of strength at $z=0$ of $B_{f,0}= 43\pm 7$ nG and an evolution with redshift with slope $\alpha =0.8\pm 0.5$; 
\item An astrophysical term whose RRM quickly decreases with the redshift. 
\end{itemize}

However, the comparison between these results and those in Ref. \citep[][]{2025A&A...693A.208C} shows that the results at low redshift depend on the actual stage of development of the astrophysical scenario in simulations, which is surely subject to improvement. 
Modern numerical simulations are attempting, with various degrees of success, to cover the giant dynamical range needed to resolve at the same time the growth of cosmic structures on ($\geq 10-100 ~\rm Mpc$) scales, as well as the process of star formation and feedback from active galactic nuclei, which ultimately depends on physical processes on $\leq\rm pc$ scales. This quest ultimately requires the adoption of sub-grid recipes to account for unresolved processes, with still large uncertainties and outcomes (e.g., \cite{2020NatRP...2...42V,2022MNRAS.511.3751H,2023Galax..11...73B, 2025arXiv250206954V,2025ApJ...980...61M}). While the simulations used here represent a significant advancement in combining magnetic fields, cosmic rays, and galaxy evolution \cite{va25a}, they clearly still are limited by resolution and cannot represent the final stage in the simulation of the astrophysical seeding of magnetic fields. 

Therefore, one can also give the results independently of the magnetogenesis model and of the astrophysical contribution shape, which returns a strength in the range $B_{f,0}= 10$--60 nG, and a redshift evolution spanning from an invariant proper field to an invariant comoving field. Regardless, both works favor
a dominant primordial magnetic field, which thus is our most important result of the RRM versus redshift analysis.

\textls[-15]{We have also presented consistency checks between simulations and observed data of stacking synchrotron emission from cosmic filaments at low frequency by Ref. \cite{2021MNRAS.505.4178V,vern23} in Section \ref{stacking}. 
When these results are combined with RRM rms data, we get a consistent picture of 
magnetic fields in the cosmic web: also here, the combined analysis suggests the presence of magnetic fields of several tens of nanoGauss in the $z \leq 3$ cosmic web, and it 
excludes a sole astrophysical scenario for the origin of these fields.
Between many possible primordial scenarios, our joint analysis
prefers either inflationary-like ($P_B \propto k^{-1}$), or causal ($P_B \propto k^2$) scenarios, with initial field amplitude $\leq 0.3 \rm ~nG$.
The causal scenario, however, is unfavored, as it can reproduce at the same time synchrotron stacking and RRM rms only if the initial 
amplitude of the field is set $\sim 10$--15 times higher than allowed by 
CMB limits. As a caveat, we notice that causal fields are the ones that are most difficult to be simulated in finite difference MHD methods, as energy fluctuations concentrated on scales close to the numerical resolution can be progressively smeared out by artificial dissipation. Hence, another line of future numerical investigation will focus on an analysis of this challenging case.}

A third investigated probe, that is the RRM radial profile moving away from galaxy groups, recently detected with POSSUM data \citep[][]{2024MNRAS.533.4068A}, was simulated as well, but it leads to inconclusive results (Section \ref{askap}). The RRMs predicted by the simulations of the primordial stochastic $nB =-1$ scenario combined with the astrophysical model are larger and closer to the observed values than those predicted for the sole astrophysical scenario. This points to favor the presence of a dominant primordial magnetic field component, as for the other two probes. However, MAD RRMs fall short of the observed values by a factor of $\approx 2$, indicating that the simulations are not yet complete, as discussed in Section \ref{askap}, and makes these results not yet conclusive.

We stress that a hypothetical detection of a causal field with normalization in the range $\sim 0.01-0.1 \rm ~nG$ would be particularly interesting, as this magnetic field model can also potentially generate the stochastic gravitational wave background observed by Pulsar Timing Arrays \cite[][]{PTA}, provided that this happened during the QCD phase transition (i.e., energy scale of about $\sim 150 \rm ~MeV$) around $z \sim 10^{12}$, i.e., during the first $\sim ~\mu s$ of cosmic evolution (e.g., \cite{2012ApJ...759...54T,2021PhRvD.103d1302N}. However, the exact dynamical evolution of such primordial fields from generation to recombination epochs is theoretically uncertain, owing to the little known details of their inverse cascade and the role of local and global helicity (e.g., \cite{2023NatCo..14.7523H,2024A&A...687A.186B}. Nevertheless, the possibility of such a striking connection between the magnetization properties of the low-redshift Universe observable with radio telescopes, and the primordial Universe in an epoch which is impossible to probe through electromagnetic radiation is extremely fascinating and surely deserves more investigation. 

\section{Conclusions}
\label{sec:conc}

In this work, we compared radio observables with the predictions of the most updated MHD cosmological simulations of the cosmic web to check which scenarios for the origin of cosmic magnetism are favored by the data. The radio probes we used are RRM rms at low frequency in filaments as a function of the redshift, stacking of synchrotron emission in filaments at low frequency, and MAD RRM radial profile just off the galaxy group. 

We found that the first two of them favor the presence of a dominant primordial magnetic field component and disfavor a sole astrophysical scenario, and the third probe does not yet give an unambiguous outcome. 

We also estimated the average field strength in filaments. Independently of the scenario and the shape of the RRM astrophysical component, it is in the range of 10--60 nG at $z=0$, with a redshift evolution ranging from an invariant proper field to an invariant comoving field. When restricted to the model that gives the best match to the simulations (a primordial stochastic $nB=-1$ scenario combined with the astrophysical model), it gives $43\pm 7$ nG and a slope of $\alpha=0.8\pm 0.5$, with an astrophysical component RM rapidly decreasing with the redshift. 

This result is in contrast with the results of Ref. \citep[][]{2025A&A...693A.208C}, where, based on a less-developed cosmological model, we found that an RRM astrophysical component increasing with redshift was needed. This suggests that uncertainties still exist in constraining the redshift evolution of the astrophysical components of RRM, which can be likely fixed with further developments of the astrophysical model. A further and much-needed improvement is an independent knowledge of the RRM astrophysical component shape, which is best obtained with surveys such as POSSUM that observe at gigahertz frequencies, where this component dominates over that of the cosmic web. 

The circumstance that radio observations alone can constrain primordial magnetic fields 
several times deeper than what is doable with present CMB analysis (e.g., \cite{2019JCAP...11..028P}) makes the upcoming advent of the square kilometer array (SKA) particularly exciting.
Simulations predict that direct imaging of at least the ``tip of the iceberg'' of radio emission from shocks in filaments might be achieved by the SKA \citep[][]{va15ska,va15radio}, provided that the stacked radio flux detected by Ref. \cite{2021MNRAS.505.4178V,vern23} is contributed by a fraction of bright objects. In principle, the SKA-LOW should be best instrument for this, owing its sensitivity to large-scale emission ($\sim 1^\circ$), yet the SKA-MID might allow performing increasingly deeper stacking studies, considering the high degree of polarization for the cosmic web emission reported by \cite{vern23}, also including bridges in between clusters of galaxies (e.g., \cite{2024A&A...691A.334V}).


While collecting at the same time more RM data from the Milky Way foreground, POSSUM and the SKAO will allow a more firm reconstruction of the extragalactic RRM rms contribution (e.g., \cite{2017ARA&A..55..111H}), and it will enable to use the complementary information by fast radio bursts, whose simultaneous detection of dispersion and rotation measurements will allow modeling the free-electron distribution along the line of sight, and hence the parallel magnetic field component (e.g., \cite{2016ApJ...824..105A,hack19,2023ApJ...954..179M}). 

At the same time, SKAO's planned science activities on cosmic magnetism will be complemented by the exploration of 
magnetic fields in voids by the new Cherenkov telescope array (\href{https://www.cta-observatory.org/}{CTA}), currently under construction \cite{2011ExA....32..193A}.
The CTA will use several sub-arrays of telescopes, to extend the dynamic energy range of observations down to $\sim 10$~GeV and up to $\sim 100$~TeV, and this should allow increasing the existing lower limits from ICC studies (or obtain detections) on the level of $\sim 10^{-12} \rm G$ \cite{2019scta.book.....C}.

In closing, assessing the origin of cosmic magnetism, its impact on the evolution of the cosmic web, and its links with fundamental processes in the early Universe is a key astrophysical endeavor, to which radio astronomy is, and will be in the future, providing vital milestones.

\authorcontributions{Formal analysis, E.C. and F.V.; Investigation, E.C. and F.V.; Methodology, E.C. and F.V.; Resources, E.C. and F.V.; Software, E.C. and F.V.; Validation, E.C. and F.V.; Writing---original draft, E.C. and F.V.; Writing---review and editing, E.C. and F.V. All authors have read and agreed to the published version of the manuscript.}

\funding{ {F.V.'s} 
research has been supported by Fondazione Cariplo and Fondazione CDP, through grant no. Rif: 2022-2088 CUP J33C22004310003 for the ``BREAKTHRU'' project.
F.V. also acknowledges financial support from the European Union's Horizon 2020 program under the ERC Starting Grant ``MAGCOW'', no. 714196.}


\dataavailability{The 190 LOS of the models astrophysical and combined are available at the web site `The Radio Cosmic Web' at the following link: 
\url{https://cosmosimfrazza.eu/radio-web} (accessed on 15/05/2025).}

\acknowledgments{F.V. acknowledges the CINECA award ``IsB28\_RADGALEO'' and ``IsCc4\_\linebreak FINRADG'' under the ISCRA initiative, for the availability of high-performance computing resources and support.}

\conflictsofinterest{The authors have no conflicts of interest}


\begin{adjustwidth}{-\extralength}{0cm}

\reftitle{References}



\PublishersNote{}
\end{adjustwidth}
\end{document}